\documentclass{article}

\usepackage{./tabularray}
\usepackage{graphicx}
\usepackage{booktabs}
\usepackage{amssymb}
\usepackage{caption}
\usepackage{graphicx}
\usepackage{float} 
\usepackage{amsmath}
\usepackage{subfigure}
\usepackage[font={small}]{caption}
\usepackage{color}
\usepackage{wrapfig}
\usepackage{spconf,graphicx}

\usepackage{enumitem}
\setlist{nosep, leftmargin=14pt}

\usepackage{mwe} 

\title{Author guidelines for ISBI proceedings manuscripts}
\title{Rethinking Dual-Domain Undersampled MRI reconstruction: domain-specific design from the perspective of the receptive field}

\name{Ziqi Gao$^{\star \dagger}$ \qquad S. Kevin Zhou$^{\star \dagger}$ }

\address{$^{\star}$ School of Biomedical Engineering, Division of Life Sciences and Medicine,\\University of Science and Technology of China, Hefei, Anhui, China 230026 \\ $^{\dagger}$ Center for Medical Imaging, Robotics, Analytic Computing \& Learning (MIRACLE), \\ Suzhou Institute for Advance Research,\\ University of Science and Technology of China, Suzhou, Jiangsu, China 215123}

\begin{document}
%
\maketitle
\begin{abstract}
Undersampled MRI reconstruction is crucial for accelerating clinical scanning. Dual-domain reconstruction network is performant among SoTA deep learning methods. In this paper, we rethink dual-domain model design from the perspective of the receptive field, which is needed for image recovery and K-space interpolation problems. Further, we introduce domain-specific modules for dual-domain reconstruction, namely k-space global initialization and image-domain parallel local detail enhancement. We evaluate our modules by translating a SoTA method DuDoRNet under different conventions of MRI reconstruction including image-domain, dual-domain, and reference-guided reconstruction on the public IXI dataset. Our model DuDoRNet+ achieves significant improvements over competing deep learning methods.
\end{abstract}
\section{Introduction}
\label{sec:intro}
Magnetic resonance imaging (MRI) is a non-invasive and flexible imaging modality widely used in clinical practice. 
Complete K-space measurements lead to unbearable acquisition time while fewer measurements lead to aliasing and blurring in the image. Under-sampled MRI reconstruction aims to reconstruct the high-quality, clean MRI image from its low-quality, aliased counterpart. Previously, Compressed Sensing (CS) and Parallel Imaging (PI) accelerated MRI reconstruction 2-3 times. Since the revolutionary work~\cite{wang2016accelerating}, convolutional neural networks (CNN) have become the primary workhorse for under-sampled MRI reconstruction. As the success of Transformer~\cite{dosovitskiy2020vit} is now indisputable in computer vision, Transformer has shown great potential for undersampled MRI reconstruction as well~\cite{huang2022swinmr,zhou2023dsformer,guo2022reconformer,feng2021task,lyu2022dudocaf}.

Many methods ~\cite{yang2017dagan,jin2017unetmri,quan2018compressedgan} focus on adapting novel architecture designs of image-domain neural networks. Customizing conventional CNNs or ViT to MRI further benefits MRI reconstruction, including K-space data consistency (DC) ~\cite{schlemper2017dc,qin2018dc1}, dual-domain recurrent learning ~\cite{eo2018kiki,zhou2020dudornet}, over-complete representation ~\cite{guo2021oucr}. SwinMR ~\cite{huang2022swinmr} and DSFormer ~\cite{zhou2023dsformer} pioneer Swin transformer~\cite{liang2021swinir} as a strong backbone in single-contrast and multi-contrast MRI reconstruction, respectively. ReconFormer ~\cite{guo2022reconformer} develops a recurrent pyramid Transformer. Feng et al. ~\cite{feng2021task} proposed a task transformer network for joint MRI reconstruction and super-resolution.

Among all SoTA methods, dual-domain networks show strong performance by considering MRI reconstruction in both image and K-space domains. DuDoRNet ~\cite{zhou2020dudornet} presents a dual-domain~\cite{eo2018kiki} recurrent learning strategy and used a dilated residual dense CNN to eliminate non-local aliasing artifact in the image domain. DuDoCAF ~\cite{lyu2022dudocaf} leverages the query-key mechanism of MSA. Yet their models use the same networks for both domains, neglecting the distinct properties of image recovery and k-space recovery problem.

In our study, (1) we analyze the feasibility of K-space interpolation from the perceptive of the receptive field; 
(2) we propose two novel domain-specific modules, including image-domain parallel local detail enhancement and k-space global initiation; and (3) extensive experiments on the public IXI-dataset validate the superiority of our model, DuDoRNet+, under multiple settings of MRI reconstruction including image-domain, dual-domain, and dual-domain undersampled MRI reconstruction guided by a reference protocol. 

\section{Analysis and Method}
\subsection{Problem Formulation}
Let $k_u  \in \mathbb{C}^{mn} $ and $k_f  \in \mathbb{C}^{mn} $ be the under-sampled and fully-sampled k-space signal respectively; $i_u  \in \mathbb{C}^{mn} $, $i_f  \in \mathbb{C}^{mn} $, and $i_r  \in \mathbb{C}^{mn} $ be the under-sampled, fully-sampled and reconstructed image signal respectively;  $M \in \mathbb{R}^{mn} $ be the binary k-space mask for acceleration.

The under-sampled MRI reconstruction can be formulated as an image recovery problem ~\cite{wang2016accelerating,guo2021oucr,chen2020ode,lee2018mp,huang2022swinmr} with K-space DC as a regularisation term~\cite{schlemper2017dc}:
\begin{equation}
    \mathop{\arg\min}\limits_{\theta_{i}} \left(\left\|i_{f}-\mathcal{P}_{i}\left(i_{u} ; \theta_{i}\right)\right\|_{2}^{2}\right. \\
    \left.+\lambda\left\|k_{u}-M \odot \mathcal{F}\left(\mathcal{P}_{i}\left(i_{u} ; \theta_{i}\right)\right)\right\|_{2}^{2}\right),
    \setlength\belowdisplayskip{3pt}
\end{equation}
where $\mathcal{F}$ is 2D discrete Fourier Transform and the approximation function $\mathcal{P}_{x}(\cdot;\theta_x)$ is used to predict a reconstructed signal $x_r$ given its parameter $\theta_x$ and any under-sampled input.  
A few works ~\cite{zhou2020dudornet,eo2018kiki,lyu2022dudocaf} leverage the relationship between image and k-space domain using Fourier Transform pairs ($\mathcal{F}, \mathcal{F}^{-1}$) and transform MRI reconstruction into a multi-variable optimization problem described as
\begin{equation}
\begin{split}
    \mathop{\arg\min}\limits_{\theta_{i},\theta_{k}} & \left(\left\|k_{f}-\mathcal{P}_{k}\left(\mathcal{F}\left(\mathcal{P}_{i}\left(i_{u} ; \theta_{i}\right)\right) ; \theta_{k}\right)\right\|_{2}^{2} \right. \\
    & + \left\|i_{f}-\mathcal{P}_{i}\left(\mathcal{F}^{-1}\left(\mathcal{P}_{k}\left(k_{u} ; \theta_{k}\right)\right) ; \theta_{i}\right)\right\|_{2}^{2} \\
    & + \left. \lambda\left\|k_{u}-M \odot \mathcal{F}\left(\mathcal{P}_{i}\left(\mathcal{F}^{-1}\left(\mathcal{P}_{k}\left(k_{u} ; \theta_{k}\right)\right) ; \theta_{i}\right)\right)\right\|_{2}^{2}\right),
\end{split}
\end{equation}
and solve it using a dual-domain recurrent learning strategy~\cite{zhou2020dudornet}. 

Yet most of existing works design the approximation function $\mathcal{P}_{x}(\cdot;\theta_x)$ using neural networks whose architectures \textbf{mainly concentrate on the image recovery problem, neglecting the distinct properties of k-space recovery problem}. We argue that a hand-crafted design on dual-domain approximation functions is crucial for solving this multi-variable optimization problem since image and k-space bear different properties and the tasks in two spaces shall be distinct further.

\subsection{Solving K-space recovery problem from the perspective of the receptive field}
Organs scanned by MRI (e.g brains, knees) are usually rich in contextual details, where a larger receptive field can improve the image reconstruction performance in general. On the other hand, undersampled MRI reconstruction is a K-space recovery problem and has been considered as a local interpolation in many CS algorithms~\cite{griswold2002grappa}. However, when MRI is under-sampled with a higher rate, the average number of measurements in a fixed local region decreases. 

Consider a set of K-space measurements after accelerating MRI with rate $a$ using 1D Cartesian sampling pattern. Conventionally,  a central fraction of K-space measurements called auto-calibration region is fully sampled, accounting for a fraction of $R_{acs}$; while the other measurements are sampled randomly.

For the non-calibration measurements, the feasibility of interpolation depends on the number of k-space scanning lines in a reachable local region $K$ of size $k*k$, in which the minimal number of data points is two. The probability $P$ of a feasible interpolation is  
\begin{figure}
   \centering   \includegraphics[width=0.45\textwidth]{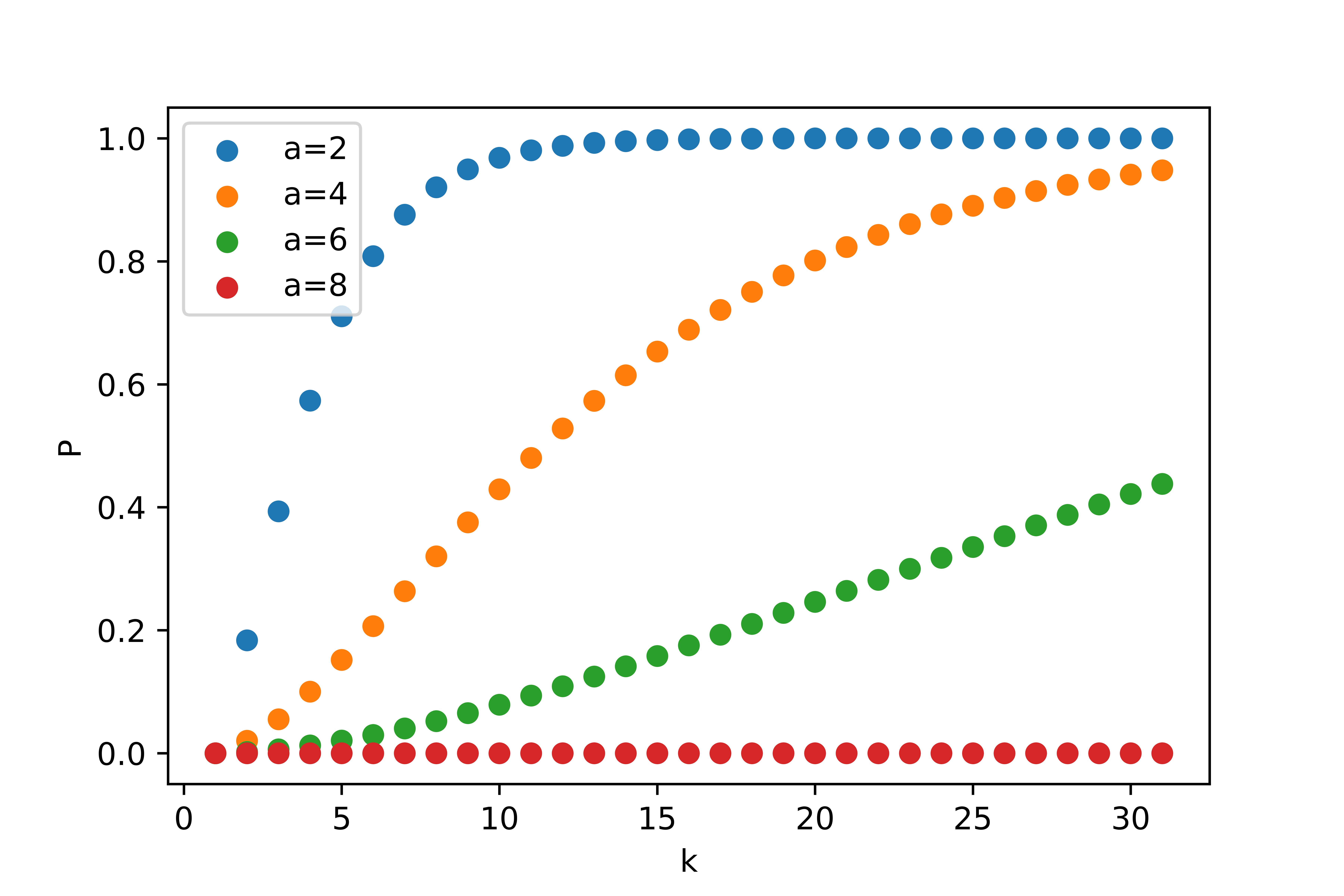}
   \caption{Probability of a feasible interpolation $P$ w.r.t. the size of receptive field and acceleration rate $a$ when $R_{acs}$=0.125.}
   \label{2}
\end{figure}
\begin{equation}
   P = 1-(p)^k-C_{1}^{k}(p)^{k-1}(1-p),
\end{equation}
where $p$ is the possibility that none of the K-space lines are measured, determined by $R_{acs}$ and $a$:
\begin{equation}
   p = 1-\frac{\frac{1}{a}-R_{acs}}{1-R_{acs}}.
\end{equation}
An illustration of the distribution of $P$ w.r.t $(K,a)$ given a fixed $R_{acs}$ is shown in figure \ref{2}. When accelerating with high rate $(>4)$, the probability of feasible interpolation is lower than $50\%$ with , not to say a successful interpolation.

Using ConvNets for MRI reconstruction ~\cite{guo2021oucr} ~\cite{zhou2020dudornet} can achieve large receptive field by upsampling, dilation convolution or large kernels. However, dilation convolution does not lead to an increase of lower bound for doing interpolation in k-space since dilation increases receptive field with aligned atrous part. Upsampling heavily burdens the GPU. Also, piling up large convolution kernels harms model convergence.

\subsection{Method}
\begin{figure}[t]
    \centering
    \includegraphics[width=0.9\linewidth]{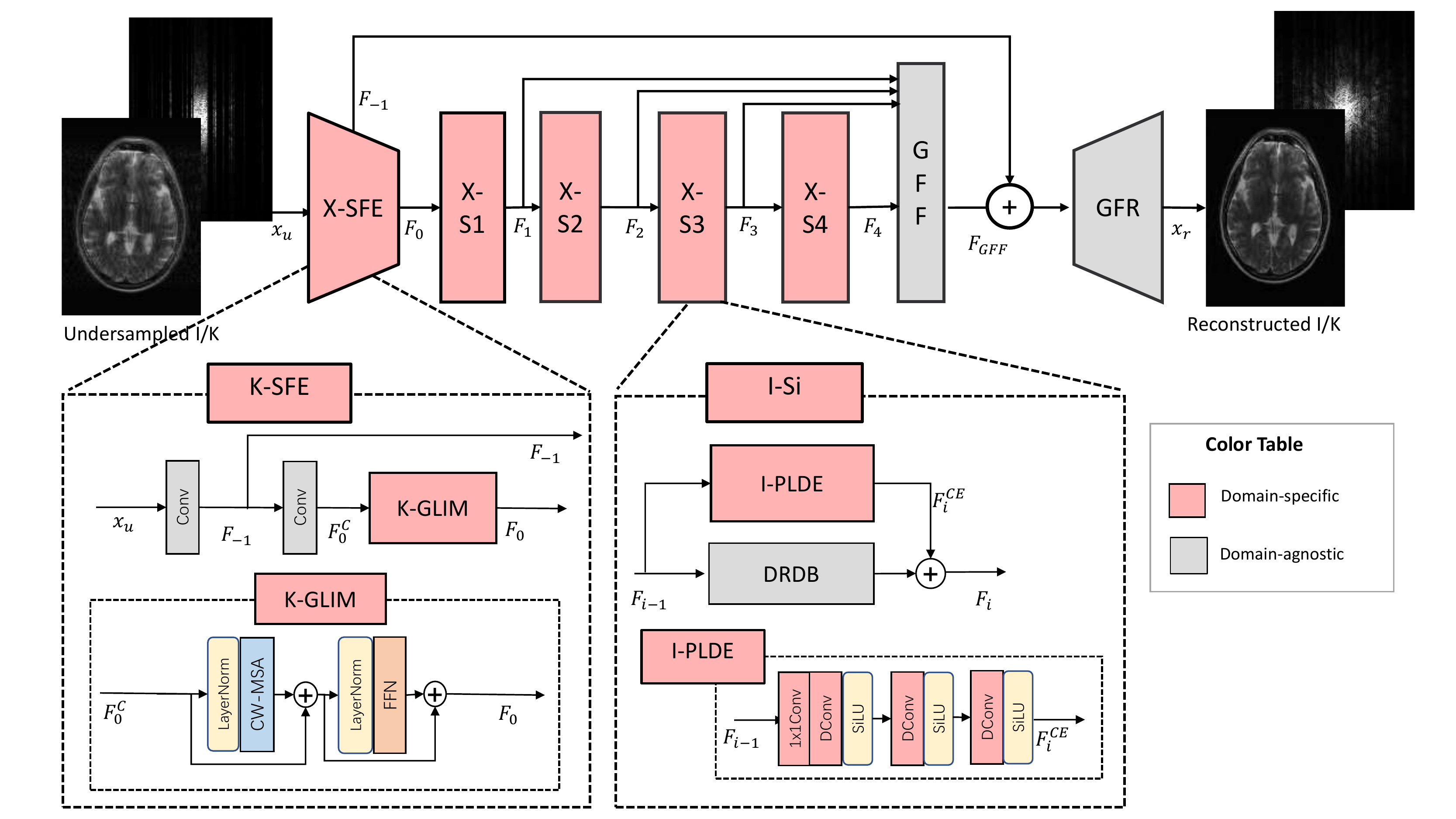}
    \caption{Framework of a recurrent block in DuDoRNet+. It is constructed by a domain-specific Shallow Feature Extraction (X-SFE), Global Feature Refinement(GFR) and 4-stage domain-specific  building blocks(X-$S_i$).  Building blocks' color follow the notation in the color table. Convolution is followed by ReLU unless noted.} 
    \label{net}
\end{figure}

We choose a SoTA dual-domain MRI reconstruction model DuDoRNet~\cite{zhou2020dudornet} for evaluating our domain-specific designs. Our designs improve its dual domain recurrent blocks by customizing global and local structures based on domain-specific properties. Our model, \textbf{DuDoRNet+}, is illustrated in Figure \ref{net}, including domain-specific Shallow Feature Extraction (X-SFE), Global Feature Refinement(GFR), and 4-stage domain-specific building blocks (X-$S_i$) as the backbone. Global residual learning and global feature fusion in DuDoRNet are preserved in ours. 

The overall pipeline goes as follows:
\begin{equation}
F_{-1} = Conv^3(x_u), F_{0}^{C} = Conv^3(F_{-1}).
\end{equation}
where $Conv^k$ denotes a convolution operation with kernel size $k*k$ and $F_{-1}$ denotes the first extracted feature used for global residual learning.  $F_{0}^{C}$ denotes the intermediate extracted feature by convolution. The second extracted feature $F_{0}$ is the input of X-$S_i$. For image domain, $F_{0} = F_{0}^{C} $.

For K-space, a Global Initiation Module (K-GLIM) is introduced based on the idea of a sketchy K-space initiation:
\begin{equation}
F_{0} = P_{K-GLIM}(F_{0}^C) = P_{K-GLIM}(Conv^3(F_{-1})).
\label{glim}
\end{equation}
Global Feature Fusion (GFF) fuses features from X-BB's and is input for global residual learning:
\begin{equation}
F_i = P_{X-S_i}(F_{i-1}),
\end{equation}
\begin{equation}
x_r = P_{GFR}(F_{-1}+P_{GFF}(concat(F_1,F_2,F_3,F_4))),    
\end{equation}
where GFF consists of 1x1 and 3x3 convolutions and GFR consists of two 3x3 convolutions, creating the refined reconstructed image/K-space $x_r$.
\subsubsection{K-GLIM: K-space global initiation}
We propose K-space global initialization module (K-GLIM) placed at the beginning of the K-space network. K-GLIM is applied on the transformed features of two convolution layers in SFE, providing channel-wise interaction and a global view.

The main part of K-GLIM is a Channel-wise Multi-head Self-attention (C-MSA). Instead of performing pixel-level or patch-level attention in a conventional spatial attention way, C-MSA is performed on the transpose of pixel-level tokens. Similar to MSA, C-MSA is an extension of Self-Attention (C-SA) in which $h$ times SA operations are done. C-SA can be expressed as
\begin{equation}
\boldsymbol{z}_{j}^c=\sum_{i} \operatorname{Softmax}\left(\frac{\boldsymbol{Q^T K}}{\sqrt{d}}\right)_{i} \boldsymbol{V}_{i, j}^T,
\end{equation}
and its computational complexity is $O(6(hw)C^2)$, linear to $(hw)$. C-MSA naturally captures global information and interactions for visual recognition tasks and complements (windowed-)spatial attention. Compared with channel-wise convolution, it is data-specific and fine-grained.

\subsubsection{I-PLDE: Image parallel local detail enhancement}
Recovering high-quality images from undersampled MRI measurements can be divided into two sub-tasks in the image domain: artifact removal and detail refinement. The former requires large receptive field due to the global property of signal aliasing, yet the latter task is much more a local task. Since the receptive field of a network enlarges with stacked convolutions, the detail refinement ability is possibly harmed with features propagating.

To this end, we propose the I-PLDE module, a parallel branch emphasizing local detail, inspired by the “divide-and-conquer” idea in ~\cite{xu2021vitae,zhang2023vitaev2}: modeling locality and long-range dependencies in parallel and fusing the features to account for both. Depthwise convolution .Different from ViTAE, we adopt stage-level parallel connection and use sequential dilated convolution (DRDB) for pyramid feature extraction instead of multiple parallel dilated convolution of different dilation rates. This design is customized to tiny models since a wide network design leads to a less stable training process and lower feature representation ability.
I-PLDE consists of a 1x1 convolution to match the hidden dimension with its parallel branch, three stacked depth-wise convolution layers, and a window embedding operation. SiLU is used for non-linear activation following the convention in ~\cite{xu2021vitae,zhang2023vitaev2}. The output of I-PLDE $F_i^{CE} $ is added with DRDN's output. ViTAE's design is not successful in our scenarios since the tiniest model it constructs has about 5 times the parameters of ours and different availability of large datasets.

\section{Experiment}
\label{sec:pagestyle}
\subsection{Dataset and training}
Our evaluation is carried out on the Multi-Contrast IXI dataset\footnote{https://brain-development.org/ixi-dataset/, CC BY-SA 3.0 license}. We use all 575 subjects with paired T2-PD and uniformly sample 14 slices from each subject volume. We split the dataset patient-wise into training, validation and testing set with a ratio of 7 : 1 : 2, corresponding to 5628 training images, 812 validation images, and 1610 test images each protocol. Images are center-cropped from 256x256 to 224x224. Code is written in Pytorch and experiments are performed using an NVIDIA GeForce RTX 3090. As for the rest, we follow the same experiment settings in DuDoRNet~\cite{zhou2020dudornet}.

\subsection{Performance evaluation}
\label{sec:typestyle}
 We compare our methods with other baseline deep learning methods in three conventions of MRI reconstruction: image-domain~\cite{ronneberger2015unet,zhou2020dudornet,feng2022multi,guo2021oucr,huang2022swinmr},  dual-domain~\cite{ronneberger2015unet,zhou2020dudornet,zhou2023dsformer} and reference-protocol-guided dual-domain reconstruction ~\cite{xiang2018ultra,feng2022multi,zhou2020dudornet,zhou2023dsformer,lyu2022dudocaf}. Considering backbone design, ~\cite{feng2022multi,xiang2018ultra} adopt Dense-Unet; ~\cite{zhou2023dsformer,lyu2022dudocaf,huang2022swinmr} share similar Swin-Transformer backbones derived from SwinIR ~\cite{liang2021swinir}. All models besides UNet ~\cite{ronneberger2015unet} have $\approx$ 1M parameters and UNet has 2M. All models are recurred twice and a DC is added at the end of each recurrent block. The recurrent time of local residual blocks in OUCR ~\cite{guo2021oucr} is set to 5, complying with their default setting; yet DCs at the end of their local recurrent structure are discarded while those at the end of their global recurrent structure are preserved for fairness.  All models are trained for 100 epochs and the hyper-parameters of each method are tuned on the validation set with test data held out for final evaluation. We consider 1D Cartesian sampling pattern with an acceleration rate ranging from 4 to 8; the center sampling fraction $R_{acs}$ is set to 0.125. Peak signal-to-noise ratio (PSNR) and structural similarity index (SSIM) are used as the quantitative evaluation metrics. 

\subsection{Results on undersampled MRI reconstruction}
\begin{table}[t]
\centering
\resizebox{0.48\textwidth}{!}{
\begin{tblr}{
  cell{1}{1} = {r=2}{},
  cell{1}{2} = {c=3}{},
  cell{1}{5} = {c=3}{},
  vline{2-3,8} = {1}{},
  vline{5} = {2}{},
  vline{1,2,5,8} = {1-17}{},
  hline{2} = {1-10}{},
  hline{1,3,4,10,14,16} = {-}{},
}

Acceleration   & 4x     &      &      & 8x    &    &      \\
    & PSNR & SSIM & MSE & PSNR & SSIM & MSE \\
Zero Padding                  & 25.17\textsuperscript{±1.80} & 80.20\textsuperscript{±6.34} & 223.51 & 24.16\textsuperscript{±1.79} & 79.90\textsuperscript{±6.37} & 282.28 \\
Unet                       & 31.73\textsuperscript{±1.98} & 95.48\textsuperscript{±2.12} & 53.03  & 26.98\textsuperscript{±2.01} & 91.43\textsuperscript{±3.75} & 157.41 \\
Dense-Unet                    & 32.49\textsuperscript{±2.23} & 96.47\textsuperscript{±1.79} & 46.10  & 27.70\textsuperscript{±1.97} & 92.16\textsuperscript{±3.50} & 132.27  \\
OUCR            & 32.52\textsuperscript{±2.26} & 96.57\textsuperscript{±1.76} & 45.90  & 27.80\textsuperscript{±2.00} & 92.33\textsuperscript{±3.44} & 129.40  \\
SwinIR                        & 32.75\textsuperscript{±2.27} & 96.69\textsuperscript{±1.71} & 43.60  & 27.78\textsuperscript{±2.08} & 92.49\textsuperscript{±3.45} & 130.76  \\
DuDoRNet\_I                   & 32.95\textsuperscript{±2.28} & 96.75\textsuperscript{±1.69} & 41.82  & 28.00\textsuperscript{±2.04} & 92.76\textsuperscript{±3.32} & 123.94  \\
Ours\_I                       & 
\textcolor{red}{33.01\textsuperscript{±1.98}} & \textcolor{red}{96.80\textsuperscript{±1.59}} & \textcolor{red}{40.36}  & \textcolor{red}{28.02\textsuperscript{±2.04}} & \textcolor{red}{92.80\textsuperscript{±3.36}} & \textcolor{red}{122.78}\\
Dual-DenseUnet                & 32.76\textsuperscript{±2.22} & 96.67\textsuperscript{±1.65} & 43.21  & 27.68\textsuperscript{±1.93} & 91.99\textsuperscript{±3.45} & 131.75  \\
Dual-SwinIR                & 33.37\textsuperscript{±2.44} & 97.09\textsuperscript{±1.57} & 38.86  & 27.73\textsuperscript{±2.02} & 92.16\textsuperscript{±3.50} & 131.70  \\
DuDoRNet(w/ ref)                  & 33.00\textsuperscript{±2.31} & 96.81\textsuperscript{±1.63} & 41.46  & 27.84\textsuperscript{±2.05} & 92.24\textsuperscript{±3.46} & 128.73 \\
Ours (w/ ref)                & \textcolor{red}{33.43\textsuperscript{±2.40}} & \textcolor{red}{97.02\textsuperscript{±1.59}} & \textcolor{red}{35.84}  & \textcolor{red}{28.13\textsuperscript{±2.10}} & \textcolor{red}{92.68\textsuperscript{±3.41}} & \textcolor{red}{125.27}  \\
DuDoRNet                      & 40.45\textsuperscript{±2.59} & 99.26\textsuperscript{±0.51} & 7.86  & 38.38\textsuperscript{±2.52} & 98.93\textsuperscript{±0.72} & 12.52  \\
Ours                  & \textcolor{red}{40.59\textsuperscript{±2.55}} & \textcolor{red}{99.29\textsuperscript{±0.50}} & \textcolor{red}{7.61}  & \textcolor{red}{38.46\textsuperscript{±2.52}} & \textcolor{red}{98.96\textsuperscript{±0.69}} & \textcolor{red}{12.00}\\ 
\end{tblr}
}
\caption{Quantitative comparison (PSNR(dB), SSIM(\%), MSE(*1e-5)) with baseline methods on the IXI-dataset. The first, second and third part correspond to image-domain reconstruction of PD, dual-domain reconstruction of PD and dual-domain reconstruction of PD with a reference protocol T2. The \textcolor{red}{best} results are marked as \textcolor{red}{red}.}
\end{table}

In Table 1, we demonstrated PD reconstruction evaluations using ×4, ×8 acceleration in three common approaches of MRI reconstruction: image-domain, dual-domain, and T2-guided dual-domain MRI reconstruction. The best results under the same setting and acceleration rate are colored with red. Our method achieves \textbf{the best performances} under all settings and acceleration rates. 

\subsection{Ablation Study}
We examine I-PLDE and K-GLIM on undersampled PD reconstruction without reference guidance. For each component, we apply it to image-domain and dual-domain subsequently. If there is a performance drop in either way, we apply it on K-space only to observe the performance. The reason for this design is that we weigh the utility of the image recovery network and the synergy between image and K-space reconstruction networks over K-space reconstruction since the image recovery network alone gives better results than K-space recovery network thus a default choice when reconstructing in a single domain. As shown in Table 2, applying them to the other domain leads to performance drops. 
\begin{table}[t]
    \label{ab1}
    \resizebox{0.45\textwidth}{!}{
    \centering
    \begin{tblr}{
          cell{1}{4} = {c=3}{},
          cell{2}{2} = {c=2}{},
          vline{2,4} = {1}{},
          vline{2-4} = {2}{},
          vline{2,4} = {3-10}{},
          hline{1,3,8} = {1-3}{},
          hline{1,2,3,4,8} = {2-10}{},
        }
                
                  4x PD &                           &                      & Domain(s) to put the module &                        &                         \\
                  & Modules  &    & Image                        & K-space                & Both                    \\
                  & K-GLIM & I-LDE & PSNR, SSIM                   & PSNR,SSIM              & PSNR, SSIM              \\
        DuDoRNet  &       &         & -                            & -                      & 33.00, 96.81           \\
        DuDoRNet w/ & \checkmark      &       & 32.89, 96.57                & \textcolor{red}{33.37, 96.87} & 33.34, 96.84           \\
        DuDoRNet w/ &       &    \checkmark   & \textcolor{red}{33.16,96.83}                & {32.76, 96.79} & 33.11, 96.83           \\
        
        Ours & \checkmark & \checkmark            & -   & -  & \textcolor{red}{33.43,97.02}           \\

    \end{tblr}
    }
    \caption{Domain-wise quantitative evaluation (PSNR, SSIM(\%)) of hybrid structure and domain-specific modules. The \textcolor{red}{best} results are marked as \textcolor{red}{red}.}
\end{table}

\section{Conclusion}
\label{sec:majhead}

We rethink dual-domain MRI reconstruction from the perspective of the receptive field and propose two domain-specific modules. By introducing domain-specific designs to a SoTA dual-domain MRI reconstruction model DuDoRNet, our model DuDoRNet+ surpasses popular deep learning methods in three common settings of MRI reconstruction. In the future, we will further validate our domain-specific design by transforming other SoTA models. Future work also includes extending our model from single-coil to multi-coil reconstruction and validating our model's utility under different sampling patterns.

\bibliographystyle{IEEEbib}
\bibliography{strings,refs}

\end{document}